\documentclass[prl,twocolumn]{revtex4}

\usepackage{amsmath,amsfonts}
\usepackage{amssymb,latexsym}
\usepackage{color}

\def\R{\mathbb{R}}

\def\ud{\mathrm{d}}
\def\ii{\mathrm{i}}

\makeatletter
\usepackage{epsfig}

\makeatother

\begin{document}

\title{More quantum centrifugal effect in rotating frame}

\author{J.-P. Gazeau}
\address{ APC, Univ Paris Diderot, Sorbonne Paris Cit\'e, 75205 Paris, France\\
Centro Brasileiro de Pesquisas F\'{i}sicas, 22290-180 - Rio de Janeiro, RJ, Brazil }
\author{T. Koide}
\address{Instituto de F\'{i}sica, Universidade Federal do Rio de Janeiro, C.P.
68528, 21941-972, Rio de Janeiro, RJ, Brazil}
\author{R. Murenzi}
\address{Division of Science Policy and Capacity Building, UNESCO, 7/9 place de Fontenoy 75007 Paris, France}
\begin{abstract}
The  behaviour of quantum systems in non-inertial frames is revisited from the point of view of affine coherent state (ACS) quantization. 
We restrict our approach to the one-particle dynamics confined in a rotating plane about a fixed axis. 
This plane is considered as punctured due to the existence of the rotation center, 
which is viewed as a singularity. The corresponding phase space is the affine group of the plane and 
the ACS quantization enables us to quantize the system by respecting the affine symmetry of the true phase space.
Our formulation predicts the appearance of an additional quantum centrifugal term, besides the usual angular momentum one, which prevents the particle to reach the singular rotation center.  
Moreover it  helps us to understand why two different non-inertial Schr\"{o}dinger equations are obtained in previous works.
The validity of our equation can be confirmed experimentally by  observing the harmonic oscillator  bound states   and the critical angular velocity for their existence.
\end{abstract}

\maketitle

{\it Introduction.---}
It is known that fictitious forces in non-inertial frames affect the behaviour of quantum systems, see for instance \cite{schmupleb77}. 
In 1979, behaviour of neutrons in a rotating frame is experimentally measured and 
an interference of the wave function is observed \cite{experiment}.
Sakurai pointed out that this can be understood by noting 
the analogy between the Coriolis force and the Lorentz force \cite{sakurai}. 
Indeed, the one-particle classical Hamiltonian in a rotating frame is given by 
\begin{equation}
H = \frac{1}{2m}(\vec{p} + m \vec{x} \times \vec{\Omega})^2 - \frac{m}{2} \vec{\Omega}^2 \vec{x}^2 +V(\vec{x})\, , 
\label{eqn:cla-hami}
\end{equation}
where $m$, $\vec{\Omega}$ and $V(\vec{x})$ are the mass of the particle, the angular velocity of the rotating frame 
and the potential energy, respectively.
One can see that the non-inertial effects 
are represented as  shifts of momentum and potential, as it holds in  the case of gauge interactions.

Afterwards, various approaches were proposed to derive the corresponding modified Schr\"{o}dinger equation, but  most of the derivations are of phenomenological nature \cite{mashhoon,anandan,anandan-suzuki}.
In the different approach of Ref.\;\cite{takagi1}, the concept of the Galilean covariance is extended. 
In Ref.\ \cite{koide}, a stochastic generalization of the variational calculus is applied. 
Whatever the differences between these approaches, the obtained results are compatible.

More recently, another non-inertial Schr\"{o}dinger equation, based on cocycle Galileo group representation theory, 
has been presented by Klink and Wickramasekara \cite{klink-prl}. 
In this approach, non-inertial frames are represented as the \textit{Galilei line group} and the 
quantized Hamiltonian is defined as the generator of time translations in this group. 
The derived equation, however, has a new term associated with the acceleration of the rotating frame 
and different coefficients from the equation obtained in the previous approaches.

Nevertheless, none of the above mentioned works take really into account the fact 
that, for example, the two-dimensional configuration space has a singular point at which 
the rotation axis intersects with the plane. 
At the singular point, we cannot consider a finite angular velocity and thus quantum observables 
are not well-defined there.
Such a modification of the geometry can raise difficulties, like the non-existence of essential self-adjointness \cite{reedsimon75}, 
on the level of quantum observables which are issued from a quantization based on the (canonical) Weyl-Heisenberg 
or more sophisticated Galilean symmetry of the classical phase space. 
Since the existence of the singular point breaks this symmetry, 
it is reasonable to use an alternative quantization which respects the affine symmetry of the true phase space, namely the geometry $\R^2_{\ast}\times \R^2$, where the notation $\R^2_{\ast}$ stands for the configuration punctured plane. 
Precisely, this manifold is identified with the $2D$ affine group SIM(2), and the relevant quantization 
is based on the resolution of the identity by SIM(2) coherent states, i.e. states obtained from the action on a ``fiducial vector" of a unitary irreducible representation (UIR) of SIM(2). 
This is called affine coherent state (ACS) quantization, whose the formulation is developed below.  
More general affine covariant integral quantizations of the motion on the half-line 
and punctured plane are discussed in Ref.\ \cite{gamu16-17}.

Accordingly, we study in this work the non-inertial quantum dynamics by emphasizing the role 
of symmetry resulting from the point singularity, by applying  
the ACS quantization to a particle moving in the rotating plane.  For the sake of manageable illustration, we specify the model \eqref{eqn:cla-hami} with an isotropic harmonic potential. 
Then we find that quantum fluctuations qualitatively modify the behavior of the centrifugal force and the value of 
the critical angular velocity of the rotating frame which characterizes the maximum angular velocity 
to have bound states.
The former, the additional quantum centrifugal term, can be 
experimentally verified by observing the energy spectrum of the bound states.
Moreover, our formulation helps us to understand why we had two different non-inertial Schr\"{o}dinger equations, 
Refs.\ \cite{schmupleb77,mashhoon,anandan,anandan-suzuki,takagi1,koide} and Ref.\ \cite{klink-prl}.
In the following, natural units $\hbar=1=c$ are used.

{\it Quantization with affine coherent states.---}
An affine transformation of the plane is denoted by $(a,\theta, \vec{b})$,
where $a>0$, $0\leq \theta< 2\pi$ and $\vec{b}\in \R^2$ represent the parameters of dilatation, rotation, and translation, respectively.
It acts on a vector $\vec{r}$ as 
\begin{eqnarray}
(a,\theta,\vec{b})\vec{r} = aR(\theta) \vec{r} + \vec{b}\,,
\end{eqnarray}
where $R(\theta)$ is the two-dimensional rotation matrix, 
\begin{equation}
\label{rotmat}
R (\theta) = \begin{pmatrix}
  \cos \theta & -\sin \theta \\
\sin \theta & \cos \theta
\end{pmatrix} \, . 
\end{equation}
Viewing $(a,\theta)$ as polar coordinates in the punctured plane $\R^2_\ast$, the parameter space is $\R^2_\ast\times \R^2$. 
The affine group law is found through two successive actions of the affine  transformations as 
\begin{equation}
\label{2daffine}
(a,\theta,\vec{b})(a^{\prime},\theta^{\prime},\vec{b^{\prime}}) \vec{r} 
= (aa^{\prime}, \theta + \theta^{\prime}, aR(\theta)\vec{b}^{\prime} + \vec{b}) \vec{r}\,, 
\end{equation}
with group identity $(1,0,\vec{0})$ and inverse $(a,\theta,\vec{b})^{-1}= (a^{-1},-\theta, -a^{-1}R(-\theta)\vec{b})$. 
Considering integration on the parameter space of this two dimensional $2D$-affine group SIM(2), then 
the left invariant measure is given by 
\begin{equation}
\label{dLg}
\ud_L g = \frac{1}{a^3} \ud a \ud\theta \ud^2 \vec{b}\,.
\end{equation}

Let us consider the motion of a particle with  mass $m$ in  a  plane in rotation about the origin of an inertial plane.
The phase-space variables for this motion are given by the position $\vec{q}$ and momentum $\vec{p}$, which are defined in the domains  
$ \vec{q} \in \R^2_\ast $ and $\vec{p} \in \R^2$, respectively. Since the point $\vec{q}=0$ is 
deleted from the domain of definition of the position in the rotating frame, 
these phase-space variables can be identified with the above parameters of SIM(2) as
\begin{equation}
\label{ }
|\vec{q}|\equiv q = a^{-1}, \ \ \ \vec{q} = (q\cos \theta, q\sin \theta), \ \ \ \vec{p} = \vec{b}\,.
\end{equation}   
Consequently, the canonical phase space measure  is the left invariant measure \eqref{dLg} on SIM(2),
$\ud^2\vec{q} \,\ud^2 \vec{p} = \ud_L g$.

The $2D$-affine group SIM(2) described with the above $(\vec{q},\vec{p})$ parameters 
has a UIR defined in the Hilbert space $\mathcal{H}:=L^2 (\R^2_{\ast}, \ud^2 \vec{x})$ as
\begin{equation}
\label{affUIR}
\psi (\vec{x})  \mapsto (U(\vec{q}, \vec{p}) \psi) (\vec{x}) 
= 
\frac{e^{\ii\vec{p}\cdot \vec{x}}}{q}  \psi \left(R(-\theta) \frac{\vec{x}}{q} \right)\,  .
\end{equation}
It allows to build the family of states $| \vec{q}, \vec{p} \rangle_\psi$ as
\begin{equation}
\langle \vec{x} | \vec{q}, \vec{p} \rangle_\psi 
\equiv 
(U(\vec{q}, \vec{p}) \psi) (\vec{x})\, .
\end{equation}
The properties of $U$ entail the \textit{overcompleteness} of these states, i.e, they solve the identity in $\mathcal{H}$,
\begin{equation}
\label{resunit}
\frac{1}{4\pi^2 c_\psi }\int_{\mathrm{SIM}} \ud^2 \vec{q}\, \ud^2 \vec{p}\,| \vec{q}, \vec{p} \rangle_\psi {}_\psi\langle \vec{q}, \vec{p} |  
= I\,  ,
\end{equation}
provided that $\psi$ is \textit{admissible} in the following sense
\begin{equation}
c_\psi 
:= \int^{ 2\pi}_0 \ud\theta \int^\infty_0 \frac{\ud q}{q}
 \left| \psi \left( \vec{q}\right) \right|^2 < \infty\,. 
\end{equation}
Note that if we perform the dilation $q \mapsto \kappa q$ with $\kappa >0$, the resolution of the identity 
still holds with the ``$\kappa$-states'' $| \kappa \vec{q}, \vec{p} \rangle_\psi$ and the same $c_\psi $.
The states $| \vec{q}, \vec{p} \rangle_\psi$ are named affine coherent states.

In the ACS quantization, the operator $\hat{A}_f$ corresponding to a classical observable 
or c-number $ f(\vec{q}, \vec{p})$ is defined by
\begin{equation}
\label{ACSq}
\hat{A}_f \equiv 
\frac{1}{4\pi^2 c_\psi}\int_G \ud^2 \vec{q} \,\ud^2 \vec{p}\, f(\vec{q}, \vec{p})| \vec{q}, \vec{p} \rangle_\psi {}_\psi\langle \vec{q}, \vec{p} |\, .  
\end{equation}
If $f$ is real-valued, the map $f \mapsto \hat{A}_f$ yields a symmetric operator, and if $f$ is semi-bounded, $\hat{A}_f$ is essentially self-adjoint. 
The strongest argument in favour of our quantization scheme holds in the fact that it respects the affine symmetry of the phase space, 
i.e.,  it is covariant in the following sense
\begin{equation}
\label{affcov}
U(\vec{q}, \vec{p})\,A_f\,U^{\dag}(\vec{q}, \vec{p})= A_{\mathcal{U}(\vec{q}, \vec{p})f}\, , 
\end{equation}
where $\left(\mathcal{U}(\vec{q}, \vec{p})f\right)(\vec{q^{\prime}}, \vec{p^{\prime}}):= f\left((\vec{q}, \vec{p})^{-1}(\vec{q^{\prime}}, 
\vec{p^{\prime}})\right)$ represents the classical affine symmetry as which the choice of origin $((1,0),\vec{0})$ 
in the phase space is arbitrary. Moreover, the possibility of working in 
Eq.\ \eqref{ACSq} with  the $\kappa$-states, or taking profit of the arbitrariness of the fiducial vector 
or ``wavelet'' $\psi$ with finite $c_\psi$ 
allows to adjust the free scaling parameters appearing in the expression of  $A_f$ in order to match experimental observations.

{\it Three fundamental quantization formulae.---}
Our purpose is the quantization of the standard classical Hamiltonian of the form 
$ \vec{p}^{\,2} +  \vec{p}\cdot \vec{A}( \vec{q}) + V(\vec{q})$, like \eqref{eqn:cla-hami}, by using the above procedure. Hence a few explicit formulae are needed. 
For the sake of simplicity, we will pick real-valued fiducial vector $\psi$ with finite $c_\psi$.

The ACS quantization of $u\left(\vec{q}\right)$, which is a function only of $\vec{q}$, is the multiplication operator in $\mathcal{H}$
\begin{equation}
\label{Auq}
(I) \ \ \ \hat{A}_{u} \phi(\vec{x})= \frac{1}{c_\psi}\left(\vert \psi\vert^2 \ast_{\mathrm{aff}}u\right)(\vec{x})\, \phi (\vec{x})\,, 
\end{equation}
where the (commutative) affine convolution is used,
\begin{equation}
\begin{split}
&\left(f_1\ast_{\mathrm{aff}}f_2\right)(\vec{x}):= \\ &\int^{2\pi}_{0} \ud \phi^{\prime}\int^\infty_0 \frac{\ud x^{\prime}}{x^{\prime}} 
f_1\left( \vec{x^{\prime}}\right) \,
f_2\left(R(-\phi^{\prime}) \dfrac{\vec{x}}{x^{\prime}}\right)\,,
\end{split}
\end{equation}
with $\vec{x}^{\,\prime} = (x' \cos \phi', x' \sin \phi')$.
In particular, with the simple power isotropic case, $\hat{A}_{q^{\beta}} = d_{\beta-1} \hat q^{\beta}$, where $\hat q \phi (\vec{x})= x \phi (\vec{x})$ and 
\begin{eqnarray}
d_{s} := \int_{\R^2_*} \frac{\ud^2\vec{q} }{q^{s + 3}} 
\vert \psi (\vec{q})\vert^2.
\end{eqnarray}

In a similar fashion, the ACS quantizations of $\vec{p}$ and $\vec{p}^{\,2}$ are, respectively,  
\begin{equation}
(II) \ \ \  \label{Ap}
\hat{A}_{\vec{p}}\phi(\vec{x}) =  -\ii\nabla_{\vec{x}} \phi(\vec{x}),
\end{equation}
and 
\begin{equation}
\label{Ap2op2}
(III) \ \ \ \hat{A}_{p^2}\,\phi(\vec{x})= \left[-\nabla^2_{\vec{x}}   + \frac{\sigma}{\hat{q}^2}\right] \phi(\vec{x})\, , 
\end{equation}
where 
\begin{eqnarray}
\label{sigma}
\sigma &=& 
\frac{1}{ c_\psi}
\int_{\R^2_\ast}\ud^2 \vec{q} \left| \nabla_{\vec{q}} \psi\left( \vec{q}\right) \right|^2.
\end{eqnarray} 
We see that a repulsive term $\propto 1/\vert\vec{x}\vert^2$ is obtained with strength 
$\sigma>0$. In a semi-classical interpretation, this extra centrifugal term prevents the particle to reach the singular point, whatever the value of its angular momentum.
With a suitable choice of $\psi$, 
we can make this coefficient large enough, namely $\sigma \geq 1$ \cite{reedsimon75,kowal02,asch07}, 
to make \eqref{Ap2op2} an essentially self-adjoint kinetic operator, i.e. no boundary condition is needed.

Moreover, the ACS quantization respects the canonical commutation rule since we have, from \eqref{Auq} and \eqref{Ap}, 
$[\hat q_i,\hat p_j]\propto \ii \delta_{ij} I$, $i,j \in \{x,y\}$, 
where the proportionality constant can be adjusted to one through a suitable choice of a \underline{non-isotropic} $\vert\psi \vert$. 
As an example of such a $\psi$, consider 
\begin{equation}
\label{psifactor}
\vert\psi(\vec{q})\vert^2 = f(\theta) \,g(q), 
\end{equation}
where $f(\theta)$ and $g(q)$ are positive and real functions, satisfying 
\begin{equation*}
\int^{2\pi}_0 \ud\theta \,\sin \theta f(\theta) = 0\, \ 
\int^{2\pi}_0 \ud\theta \,\cos \theta f(\theta) > 0\, .
\end{equation*}
For instance, with $f(\theta) = \pi  (\cos \theta + 1)/2$, we have 
$c_\psi = \pi^2 \int^\infty_0 dq g(q)/q$, which imposes $g(0)=0$. We then find for the quantum position vector and the canonical commutation rule,
\begin{equation}
\label{qposvec}
\hat{A}_{\vec{q}}= {\cal N}_0 \hat{\vec{q}}\, , \quad [\hat q_i,\hat p_j]=  \ii \delta_{ij}{\cal N}_0 I
\end{equation}
and for the quantum version of the angular momentum $L = q_x p_y - q_y p_x$, 
\begin{equation}
\label{ }
(\hat{A}_{{L}} \phi)(\vec{x})= -\ii {\cal N}_0 \partial_\theta \phi(\vec{x})\, , \ \vec{x}\equiv (x,\theta) \, . 
\end{equation}
The constant $\mathcal{N}_0$ is part of the set
\begin{equation}
\begin{split}
{\cal N}_s = \frac{d_s}{2c_{\psi}} = \frac{d_s}{2d_{-1}}\,, \ \ \ \ \ 
d_s = \pi^2 \int^\infty_0 \frac{\ud q}{q^{s+2}} g(q)\, . 
\end{split}
\end{equation}
Note that we can set ${\cal N}_0 = 1$ by choosing $g$ appropriately or using $\kappa-$states. 
However, for the sake of the comparison with the result of Ref.\ \cite{klink-prl}, 
we leave it as a parameter.

For the quantization scheme of the present model following the usual Weyl-Heisenberg framework, see for instance \cite{kowal02}.

{\it ACS Quantization in rotating frame.---}
Let us illustrate our formalism with the basic situation of a rotating frame about a constant axis. 
The classical Hamiltonian for a particle is given by Eq.\eqref{eqn:cla-hami} where we choose the $z$-axis orthogonal to the rotating plane and 
$\vec{\Omega} = \Omega \vec{e}_z$. Here, $\vec{e}_z$ is the unit vector for the $z$-axis and $\Omega$ can be a function of time.
For the potential, we consider the harmonic $V(\vec{x})=m\omega^2 \vec{x}^{\,2}/2$.
The quantum version $A_H$ yielded by the above ACS quantization  governs 
the following Schr\"{o}dinger equation in the rotating frame
\begin{align}
\label{schrod}
\nonumber  \ii \partial_t \phi (\vec{x},t) =   &\left[
\frac{1}{2m} \left( - \frac{1}{c^2_\psi}\nabla^2_{\vec{x}} + \vec{A}(\vec{x}) \right)^2 +\right.\\ & + A^0 (\vec{x}) 
+ {\cal N}_1 m \omega^2  \vert\vec{x}\vert^{2}
\Biggr] \phi (\vec{x},t)\, , 
\end{align}
where
\begin{equation*}
A^0 (\vec{x})= -\frac{m}{2} \Omega^2 {\cal N}^2_0 \vert\vec{x}\vert^{2} + \frac{\sigma}{2m\vert\vec{x}\vert^2}\, , \quad 
\vec{A} (\vec{x}) = m {\cal N}_0 \vec{x} \times \vec{\Omega}\, .
\end{equation*}
In this derivation, we used Eq.\ (\ref{psifactor}).

We stress again on the fact that the appearance of the ``sigma quantum centrifugal term'',  or more simply $\sigma$-term, $\sigma/(2m\vert\vec{x}\vert^2)$
is a permanent issue of our ACS quantization since $\sigma >0$ for \textit{any} admissible and differentiable  $\psi$ such that the integral in \eqref{sigma} converges.
Such a term adds the  angular momentum repulsive one, and plays the role of a centrifugal force which pushes the particle outward from the rotation axis.  It regularizes the existing point singularity due to rotation.
In the following, we choose $\psi$ such that $c_\psi =1$.

Now, the Schr\"{o}dinger equation \eqref{schrod} has still various coefficients which depend on a given form of the affine coherent state. 
When we choose $\psi$ so that $({\cal N}_0, {\cal N}_1) = ( 1, 1/2)$, then 
the two fields $A^0$ and $\vec{A}$ reproduces the gauge field corresponding to the classical Hamiltonian (\ref{eqn:cla-hami}), 
at the exception of the $\sigma$-term.
That is, without this term, our result coincides with those of Refs.\ \cite{schmupleb77,mashhoon,anandan,anandan-suzuki,takagi1,koide}.

On the other hand, if we choose $({\cal N}_0, {\cal N}_1) = ( 2, 1/2)$ and ignore the $\sigma$-term, then 
Eq.\ \eqref{schrod} agrees with the result in Ref.\ \cite{klink-prl} neglecting the angular acceleration term.
Therefore, from the point of view of the ACS quantization, a part of the differences in the two previous non-inertial Schr\"{o}dinger equations 
is attributed to the different choices of the fiducial vector $\psi$ (or $g$ in the particular case \eqref{psifactor}).

{\it Spectrum and eigenfunctions---}
The crucial question about the validation of our result with regard to the existing approaches concerns the appearance of the $\sigma$-term, 
even in the absence of quantum angular momentum.
As a matter of fact, the existence of such a term could be confirmed experimentally. 
To be more precise with our simple model \eqref{schrod}, let us analyze its stationary states.  
It is well known that the problem is analytically solvable, and eigenfunctions are calculated as 
\begin{equation}
\label{soleq}
\begin{split}
 \phi_{n,l}(\vec{x}) &\propto L_n^{(\sqrt{\sigma+l^2})}(r^2) r^{\sqrt{\sigma+l^2}} e^{-r^2/2}\, e^{\ii (l+\varsigma)\theta}\\ 
  &(n=0,1,\cdots, \ l=0,\pm 1,\cdots, \ \varsigma\in[0,1))\,,
  \end{split}
\end{equation}
where $r=|\vec{x}|/l_0$ with $l_0 = ( 4 {\cal N}_1 M^2 \omega^2)^{-1/4}$.  
$L_n^{(\beta)}(x)$ is the associated Laguerre (or Sonine) polynomial. The parameter $\varsigma$ is the quasiperiodicity parameter introduced by Kowalski et al. in \cite{kowal02} (noted $\theta$ in their paper), 
which is allowed by the topology of the punctured plane.
Then the corresponding eigenvalues read 
\begin{equation}
\begin{split}
&E_{n,l} =\\
& \omega \sqrt{2{\cal N}_1} \left( 2n+1 + \sqrt{\sigma + (l+\varsigma)^2} 
- \frac{{\cal N}_0}{2\sqrt{{\cal N}_1}} \frac{\Omega}{\omega}l \right)\, . 
\end{split}
\label{eqn:eigenv} 
\end{equation}
Imposing time-reversal symmetry \cite{kowal02} restricts $\varsigma$ to assume two values only, namely $0$ and $1/2$. 
To simplify, let us impose here  $\varsigma=0$.
To verify the value of $\sigma$, we should observe the $l$ dependence of the energy spectrum: 
if it deviates from linear dependence on $l$,  it is an evidence of a finite $\sigma$.

{\it Quantum effect on Critical Angular Velocity.---}
Note that the above eigenvalue does not have lower bound for large enough angular velocity of the rotating frame.
This corresponds to the fact that there does not exist bound state if the angular velocity exceeds a critical value. 
In fact, the corresponding classical equations of motion are given by  
\begin{eqnarray}
M \ddot{q}_x &=& 
2M \Omega \dot{q}_y - M(\omega^2 - \Omega^2) q_x, \\
M \ddot{q}_y &=& 
-2M \Omega \dot{q}_x - M(\omega^2 - \Omega^2) q_y.
\end{eqnarray}
The first terms on the right-hand side are the Coriolis forces. 
One can see that the harmonic potential is modified by the (classical) centrifugal force and 
the magnitude $\Omega$ of the angular velocity of the rotating frame  
should be smaller than $\omega$ to have bound states. That is $\Omega_{cri} = \omega$ in the classical case.

This critical angular velocity is generally affected by quantum fluctuations. In fact, 
for Eq.\ (\ref{eqn:eigenv}) to have bounded eigenvalues for any $l$, 
the following inequality should be satisfied,  
\begin{eqnarray}
 \Omega < \frac{2\sqrt{{\cal N}_1}}{{\cal N}_0} \omega.
\end{eqnarray}
When we choose $({\cal N}_0,{\cal N}_1) = (1,1/2)$ so as to reproduce the results in Refs.\ 
\cite{mashhoon,anandan,anandan-suzuki,takagi1,koide} at the exception of the $\sigma$-term, $\Omega$ 
can be larger than the maximum value in the classical case and the critical value is given by $\Omega_{cri} = \sqrt{2}\omega$.
On the other hand, when we choose $({\cal N}_0,{\cal N}_1) = (2,1/2)$ to reproduce Ref.\ \cite{klink-prl}, 
$\Omega_{cri} = \omega/\sqrt{2}$.
Note that these critical values are independent of the $\sigma$-term 
but  
our fiducial vector $\psi$ is choosen to have  
$\sigma \geq 1$ in order to avoid the introduction of any boundary condition.

Therefore, the validity of the above two results issued from the ACS quantization 
can be verified by analyzing the critical angular velocity.

{\it Concluding remarks.---}
We have favoured the ACS quantization over the Weyl-Heisenberg-based canonical quantization in any circumstance 
where the configuration space for the classical model is the punctured plane. Our choice is motivated by the symmetry of the corresponding phase space. 
Two remarkable features arise from our approach. On one hand, the inescapable appearance of the $\sigma$-term which 
prevents the particle to reach the singular point, combined with the freedom to adjust its strength to observations of the energy spectrum of the bound states. 
On the other hand, a possible rescaling and smoothing of multiplication operators issued from classical functions on the plane, 
also combined with the freedom to adjust strengths and shapes to observations of the critical angular velocity. 
These latitudes are due to the infinite range of admissible $\psi$, and also help us to understand why we had two different non-inertial Schr\"{o}dinger equations.
Actually, the choice can be even enlarged by using all resources of affine integral quantization, 
as is explained in Ref.\ \cite{gamu16-17}. 
One should not be reluctant to take advantage of such a freedom since most of the quantum models are effective. 
Think for instance to Morse or Lennard-Jone, or others, in Molecular Physics, or effective interactions in Condensed Matter Physics. After all, the main constraint to be respected is the agreement of the quantum Hamiltonian in Eq.\ \eqref{schrod} to experimental observation. The classical Hamiltonian \eqref{eqn:cla-hami} might be viewed as a sketch for building its quantum counterpart through the map $f\mapsto \hat A_f$ in \eqref{ACSq}.

J.P. Gazeau thanks  CBPF and  CNPq for financial support and CBPF for hospitality. T. Koide acknowledges the financial support by CNPq.
The ideas and opinions expressed in this article are those of the authors and do not necessarily represent the view of UNESCO.

\end{document}